# Charge Stripper Effects on Beam Optics in 180-degree Bending Section of RISP Linac

Ji-Ho Jang[*], Hyunchang Jin, Jeong Seog Song

*RISP, Institute for Basic Science, Daejeon 305-811, Republic of Korea*

The RAON, a superconducting linear accelerator for RISP (Rare Isotope Science Project), will use a charge stripper in order to increase the charge states of the heavy ions for effective acceleration in the higher energy part of the linac. The charge stripper affects the beam qualities by scattering when the heavy ions go through the charge stripper. Moreover we have to select and accelerate proper charge states between 77+ and 81+ for uranium beam case in order to satisfy the beam power requirement at an IF (Inflight Fragmentation) target. This work focuses on the beam optics affected by the charge stripper in the 180-dgree bending section.

[*]Email: jhjang@ibs.re.kr

Fax: +82-42-878-8799

# I. INTRODUCTION

The RISP (rare-isotope science project) is developing a linac which can accelerate uranium beams to 200MeV/u with the beam power of 400 kW [1]. The linac consists of an injector and superconducting linac (SCL) as schematically shown in Figure 1. In the injector section, 10 keV/u uranium beams from an ECR (electron cyclotron resonance) ion source are accelerated to 500 keV/u by an RFQ (radio-frequency quadrupole). The SCL section consists of two parts which are connected by bending sections, 90-degree in the driver linac and 180-degree in the post-accelerator. The lower energy part of SCL includes QWR (quarter-wave resonator), HWR (half-wave resonator) cavities. The SSR (single spoke resonator) cavities will be used in the higher energy part of the superconducting linac.

In order to achieve the required beam power at IF target of 400 kW, the uranium beams with two charge states, 33+ and 34+, will be selected in a LEBT (low energy beam transport) and accelerated in the injector. The charge state increases to around 79+ through a charge stripper located after the lower energy part of SCL. The 5 charge states from 77+ to 81+ of the uranium beams are selected in the bending sections. In this work, we will focus on the 180-degree bending section which was designed to be a second-order achromatic and isochronous in order to minimize the emittance growth through the bending section [2]. In the initial phase of beam commissioning of the linac the carbon stripper will be used and replaced by a liquid lithium target.

The charge stripper can affect the beam quality in the high intensity heavy ion linac [3]. We studied charge stripper effects on beam dynamics in the 180-degree bending section of RISP linac. This work summarized the results such as particle distribution after the charge stripper, removing halo particles generated by scattering of beams with the charge stripper, the charge selection, and beam dynamics in the bending section.

# II. CHARGE STRIPPER

The thickness of the carbon stripper was determined to be 1 mg/cm² to obtain the central charge state of uranium beams becomes 79+. In the charge stripper, the particle energies are reduced and the emittances are increased by the multiple scattering of particles with the carbon material.

In order to estimate the energy decrease we used SRIM code [4] and compared the result with Bethe-Bloch equation for the stopping power [5]:

$$\left\langle -\frac{dE}{dx} \right\rangle = 4\pi N_A r_e^2 m_e c^2 z^2 \frac{Z}{A\beta^2} \left[ \frac{1}{2} \ln \frac{2m_e c^2 \beta^2 \gamma^2 T_{\max}}{I^2} - \beta^2 - \frac{C}{Z} - \frac{\delta}{2} \right], \quad (1)$$

where $4\pi N_A r_e^2 m_e c^2 / A$ = 0.307 MeV g$^{-1}$ cm² for A = 1 g mol$^{-1}$. The parameters z and Z are the atomic number of incident particle and target material. $T_{\max}$ and I represent the maximum energy transfer and the effective excitation energy, respectively. The shell correction is C and the density effect correction is δ. The detail of the parameters can be found in ref. [5].

Figure 2 shows the particle distribution depending on the kinetic energies between 17 MeV/u and 20 MeV/u of uranium beams. In this simulation we assumed that all particles have the same input energy and the incidence angle is zero which means the moving direction of particles is perpendicular to the target surface. The detailed particle distribution for the incidence energy of 18.5 MeV is given in Figure 3. The red step is the SRIM simulation result and the Gaussian fitting result is given by blue line. The center value of the kinetic energy after the charge stripper is 18.039 MeV/u and the standard deviation is 0.004 MeV/u. It means that the energy deviation is about 0.02% and small enough.

The values of the kinetic energy reduction are given in Figure 4 depending on the incidence energy. The points are the SRIM simulation results and the blue line is obtained by using the Bethe-Bloch equation of eq. (1). The difference is less than 1.6% in this kinetic energy region. Hence we can use the Bethe-Bloch formula in order to estimate the kinetic energy decrease through the carbon stripper in this energy region.

The charge distribution after the carbon stripper was obtained by using the formula in ref. [6, 7]:

$$\langle Q \rangle = \langle Q_p \rangle \left[ 1 - \exp(-12.905 + 0.2124 \times Z_p - 0.00122 \times Z_p^2) \right] g(Z_t). \quad (2)$$

It is the central charge value of the charge distribution after the charge stripper with

$$\langle Q_p \rangle = Z_p \left[1 - \exp(-83.275 \times \beta/Z_p^{0.447})\right], \quad (3)$$

and

$$g(Z_t) = 1 - 5.21 \times 10^{-3}(Z_t - 6) + 9.56 \times 10^{-5}(Z_t - 6)^2 - 5.9 \times 10^{-7}(Z_t - 6)^3 \quad (4)$$

The standard deviation of the Gaussian-type charge distribution is given by

$$Q_\sigma = \left[\langle Q_p \rangle \left(0.07535 + 0.19Y - 0.2654Y^2\right)\right]^{1/2}, \quad (5)$$

where Y = <$Q_p$>/$Z_p$. The parameters $Z_p$ and $Z_t$ represent the atomic numbers of the projectile particles and target material. This formula can be applied to $Z_p \geq 54$ and the kinetic energy greater than 1.3 MeV/u [6, 7]. Figure 5 show the charge distribution after the carbon stripper with the thickness of 1 mg/cm$^2$ and the kinetic energy of 18.039 MeV/u in uranium beams. The central value of the charge distribution is 79.15 and it means that the most probable charge state is 79. The standard deviation is 1.83 in this case. In the following analysis we used this charge distribution after the charge stripper.

### III. 180-Degree Bending Section

In order to obtain the particle distribution in phase space after the carbon stripper, we also used the SRIM code [4]. The input distribution is obtained by the TRACK code [8] with the twiss parameters and beam emittances at the entrance of the charge stripper. We used the input kinetic energy of 18.5 MeV/u before the charge stripper. Then we compared the SRIM results with the particle distributions obtained by the charge stripper routine in TRACK code because the TRACK code provides a charge stripper routine which needs the information of the fraction of each charge state and the reduction of the kinetic energy. Figure 6 shows the particle distributions with 100,000 macro-particles both in SRIM and TRACK simulations after the charge stripper. The output particles are mainly scattered in x', y', and ΔW directions in the charge stripper. The number spectrum of particle distribution in those directions are given in Figure 7. The sigma values in the Gaussian fitting of the plots are given in Table 1. We found that more halo particles are produced in SRIM simulation than TRACK in x' and y' spaces. Even though sigma values in ΔW-direction are similar between input, TRACK and SRIM results, we can observe scattered particles with lower kinetic energies both in TRACK and SRIM simulation as shown in Figure 6.

The beam dynamics calculation in the 180-degree bending section was performed by using the TRACK code [8]. The lattice parameters are same as ones in Ref. [2]. The charge state distribution is obtained by using Baron's formula as explained in the previous section. The particle distributions in phase space after charge stripper are given by the charge stripper routine in TRACK code and the SRIM simulation. The TRACK simulation with both distributions is given in Figure 8.

We used 2 slits in order to remove halo particles which are generated by scattering of projectile particles in the charge stripper. The slits with full aperture of 16 mm in both horizontal and vertical directions are located between the charge stripper and the first quadrupole magnet after the stripper. We found that the positions are efficient enough to remove halo particles. Particle distributions without and with 2 slits are given in Figure 9 and Figure 10, respectively for TRACK and SRIM, at the position of the second slit. We found that the slits can effectively remove the halo particles both in transverse and longitudinal directions.

We studied the emittance behavior through the 180-degree bending section based on the TRACK simulations given in Figure 8. The rms emittances are given in Figure 11 for particle distributions of both the TRACK charge stripper routine and the SRIM simulation. We found that the overall behavior of emittances look similar between the TRACK routine and SRIM cases after 2 slits for halo collimation. The larger rms emittances of SRIM case represent the larger scattering effect in the particle distribution by SRIM simulation than TRACK as shown in Figure 6 and Figure 7.

We also note that the collimators located after the first and second bending magnets are working well to select 5 charge states of uranium beams in order to accelerate in the higher energy part of RISP linac as shown in Figure 12. The beam losses are localized in slits of halo scrapers and collimators of charge selection systems.

## IV. CONCLUSION

We studied how the charge stripper affects the beam dynamics in the 180-degree bending section, especially focused on the carbon stripper with the thickness of 1 mg/cm$^2$. We found that the kinetic energy is reduced by about 0.5 MeV/u in the energy range between 17 MeV/u and 20 MeV/u. The halo particles generated by scattering through the charge stripper can be effectively eliminated by 2 slits located after the charge stripper. The 5 charge states of uranium beams can be selected in 2 collimators in the bending section.

## ACKNOWLEDGEMENT

This work was supported by the Rare Isotope Science Project of Institute for Basic Science funded by Ministry of Science, ICT and Future Planning.

## REREFENCES

[1] D. Jeon, et al., J. Korean Phys. Soc. **65**, 1010 (2014).

[2] Hyunchang Jin, Ji-Ho Jang, Hyojae Jang, In-Seok Hong and Dong-O Jeon, Nucl. Instrum. Methods. Phys. Res. **A795**, 65 (2015).

[3] P.N. Ostroumov, V.N. Aseev and B. Mustapha, Phy. Rev. ST Accel. Beams **7**, 090101 (2004).

[4] Website for SRIM code: http://www.srim.org/.

[5] D. Groom, Particle Data Group Notes, PDG-93-06, 8 December 1993.

[6] E. Baron, M. Bajard, and Ch. Ricaud, Nucl. Instrum. Methods. Phys. Res. **A238**, 177 (1993).

[7] A. Leon, S. Melki, D. Lisfi, J. P. Grandin, P. Jardin, M. G. Suraud, and A. Cassimi, Atomic Data

and Nucl. Data Tables, **69**, 217 (1998).

[8] P.N. Ostroumov, V.N. Aseev and B. Mustapha, TRACK, ANL Technical Note, Updated for version 3.7.

Table 1. The σ-values of Gaussian fit of the particle distribution after the charge stripper.

|  | x' [mrad] | y' [mrad] | ΔW/W [%] |
|---|---|---|---|
| Initial distribution before charge stripper | 0.551 | 0.695 | 0.068 |
| TRACK result after charge stripper | 0.609 | 0.745 | 0.071 |
| SRIM result after charge stripper | 0.684 | 0.803 | 0.073 |

Fig. 1. Layout the RISP linear accelerator.

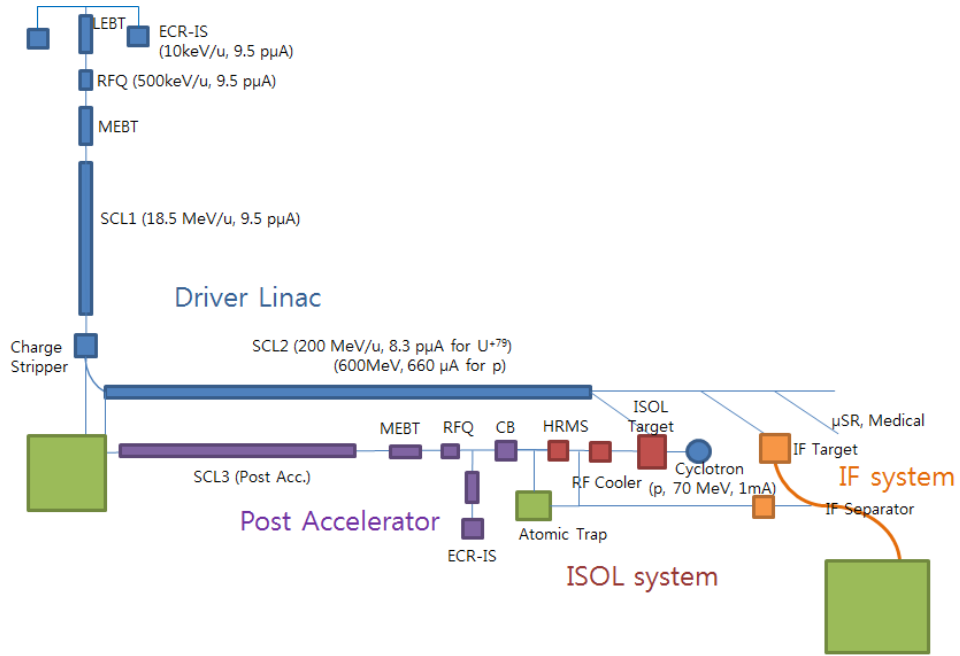

Fig. 2. Kinetic energy change and particle distribution of uranium beam depending on the incident kinetic energy.

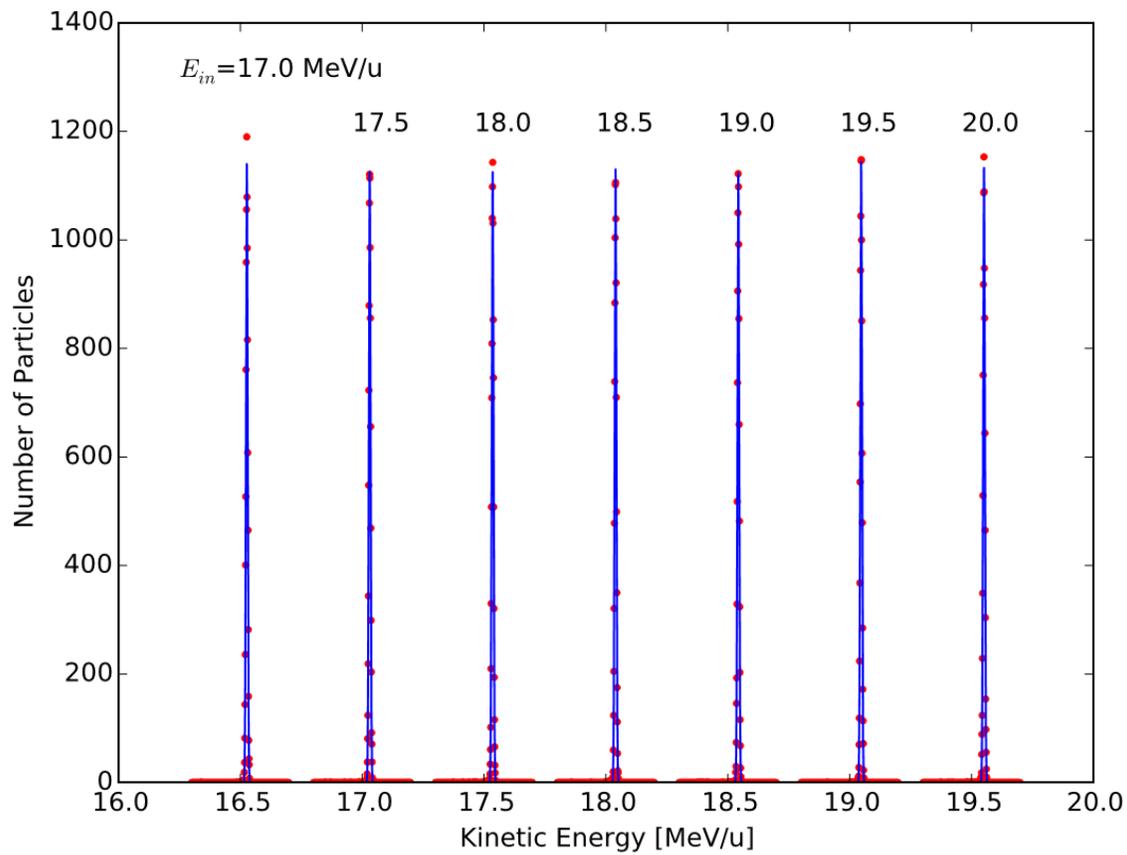

Fig. 3. Particle distribution for the incident kinetic energy of 18.5 MeV/u: red step for the SRIM simulation data and the blue line for the Gaussian fitting result.

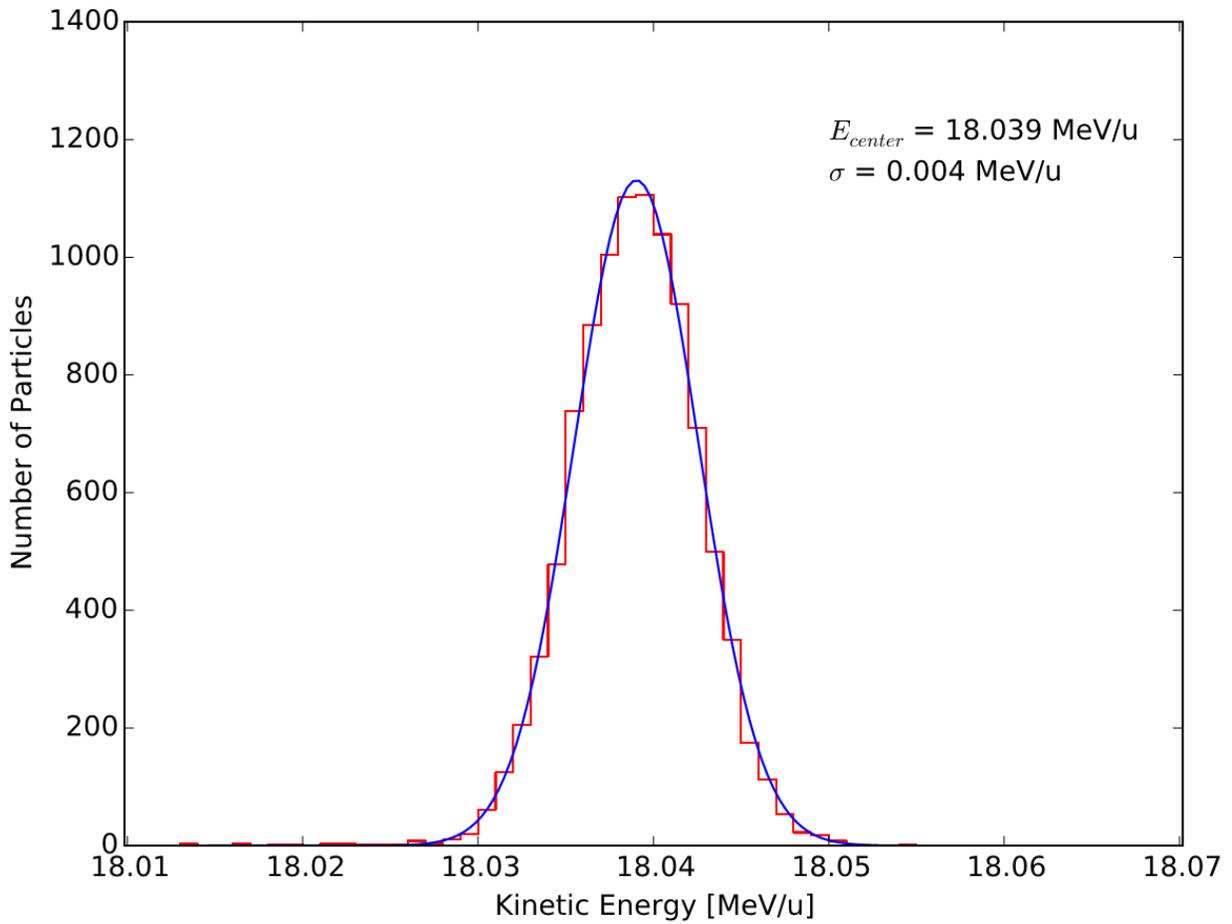

Fig. 4. Energy loss in the carbon stripper depending on the incident kinetic energies: red dots for SRIM simulation and blue line for Bethe-Bloch equation.

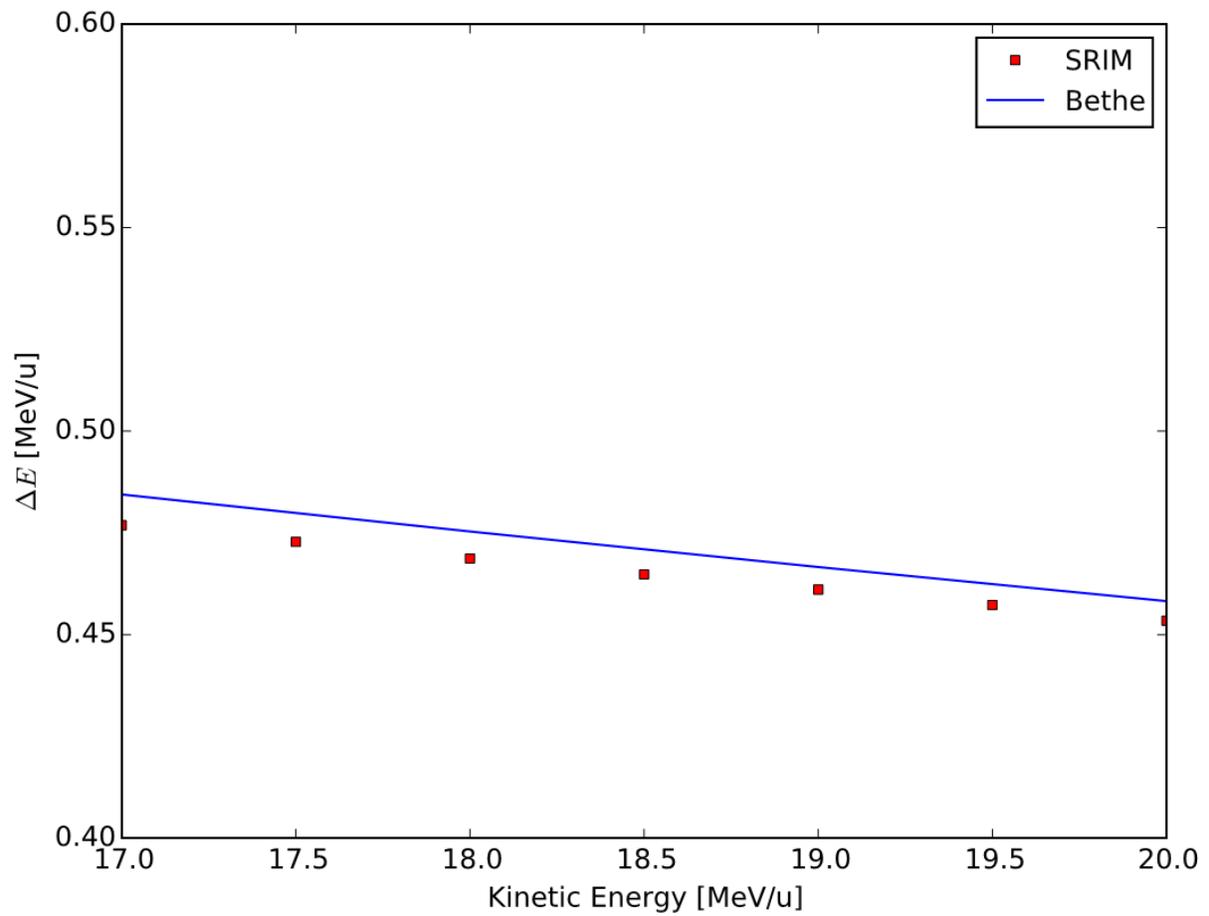

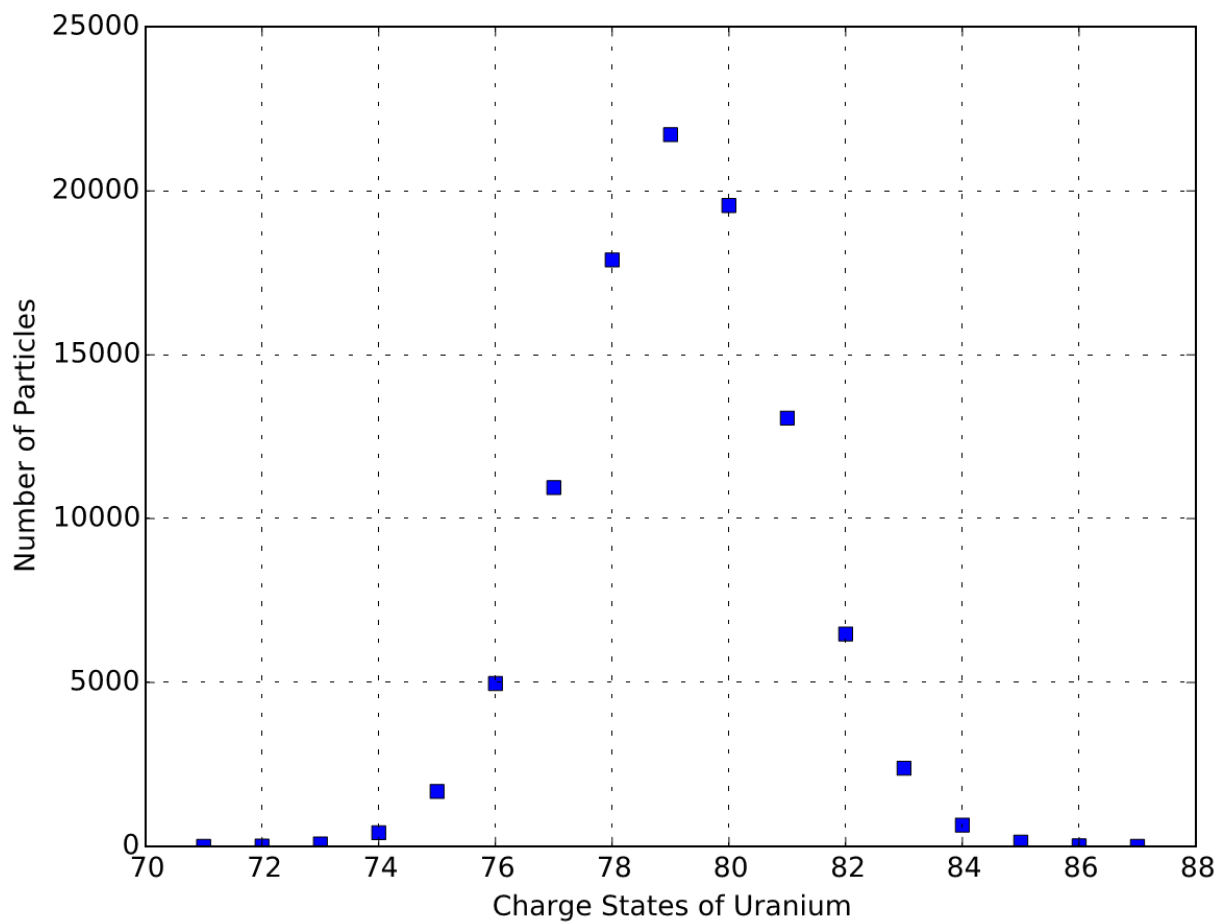

Fig. 5. Charge distribution after the charge stripper by using formula in references [5, 6].

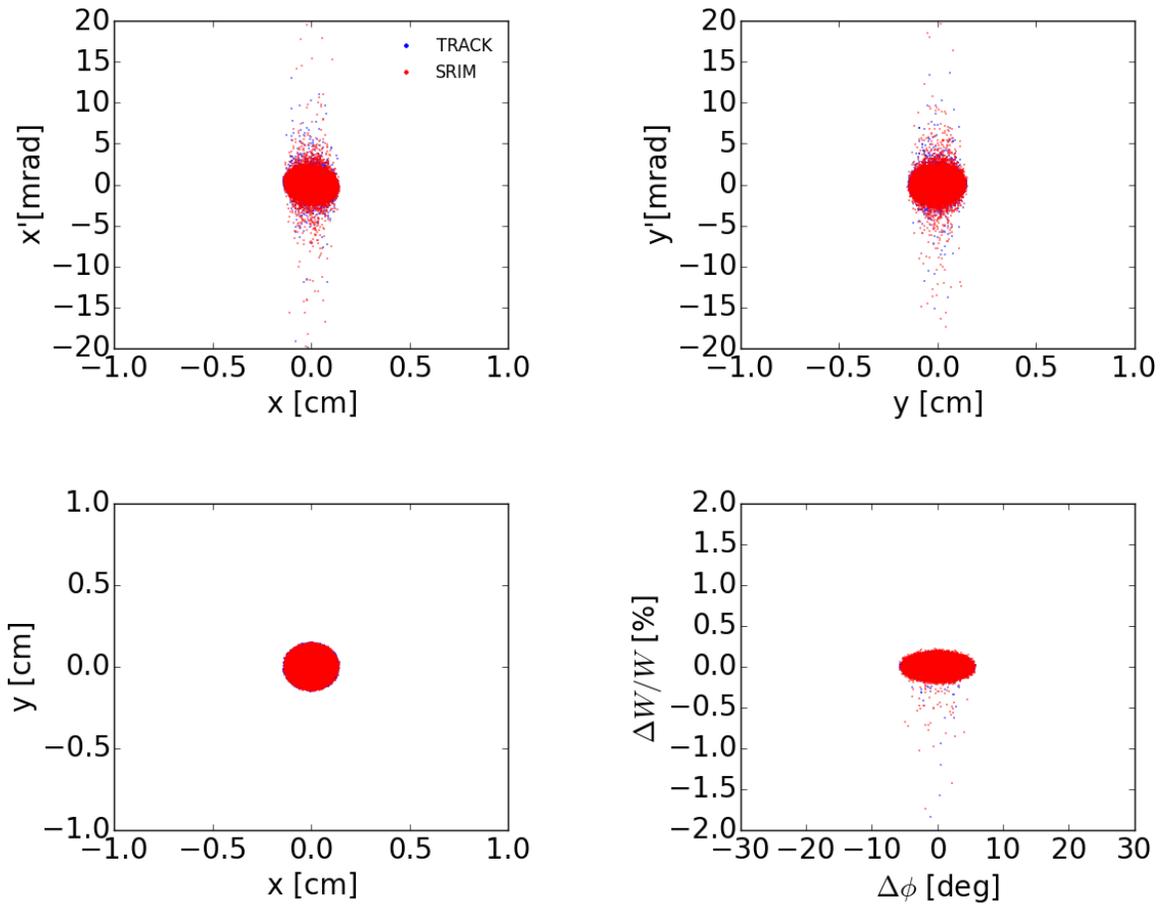

Fig. 6. Particle distributions after the charge stripper in the TRACK and SRIM simulation.

Fig. 7. Number spectrum of particle distribution: input before charge stripper, TRACK and SRIM results after charge stripper in (a) x'-direction (b) y'-direction and (c) ΔW/W-direction.

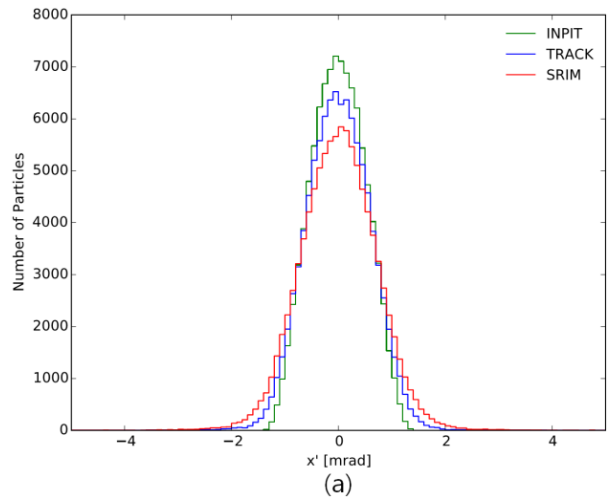

(a)

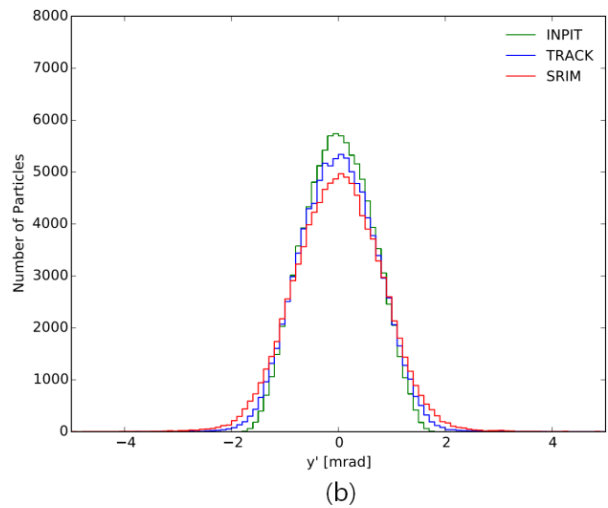

(b)

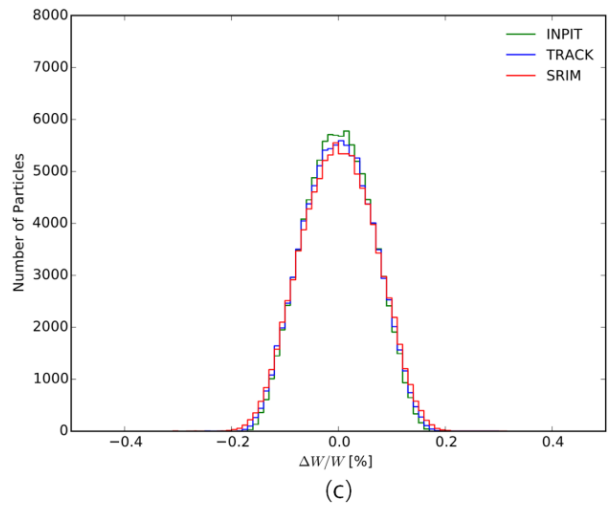

(c)

Fig. 8. TRACK simulation results through the 180-degee bending section with the particle distribution after the charge stripper obtained by (a) TRACK charge stripper routine and (b) SRIM simulation.

(a)

(b)

Fig.9. Particle distributions through slits located after the charge stripper in TRACK simulation: (a) x-x' space before and after the first silt, (a) x-x' space before and after the second silt, (c) y-y' space before and after the first silt, (d) y-y' space before and after the second silt, (e) Δϕ-ΔW space before and after the first silt and (f) Δϕ-ΔW space before and after the second silt.

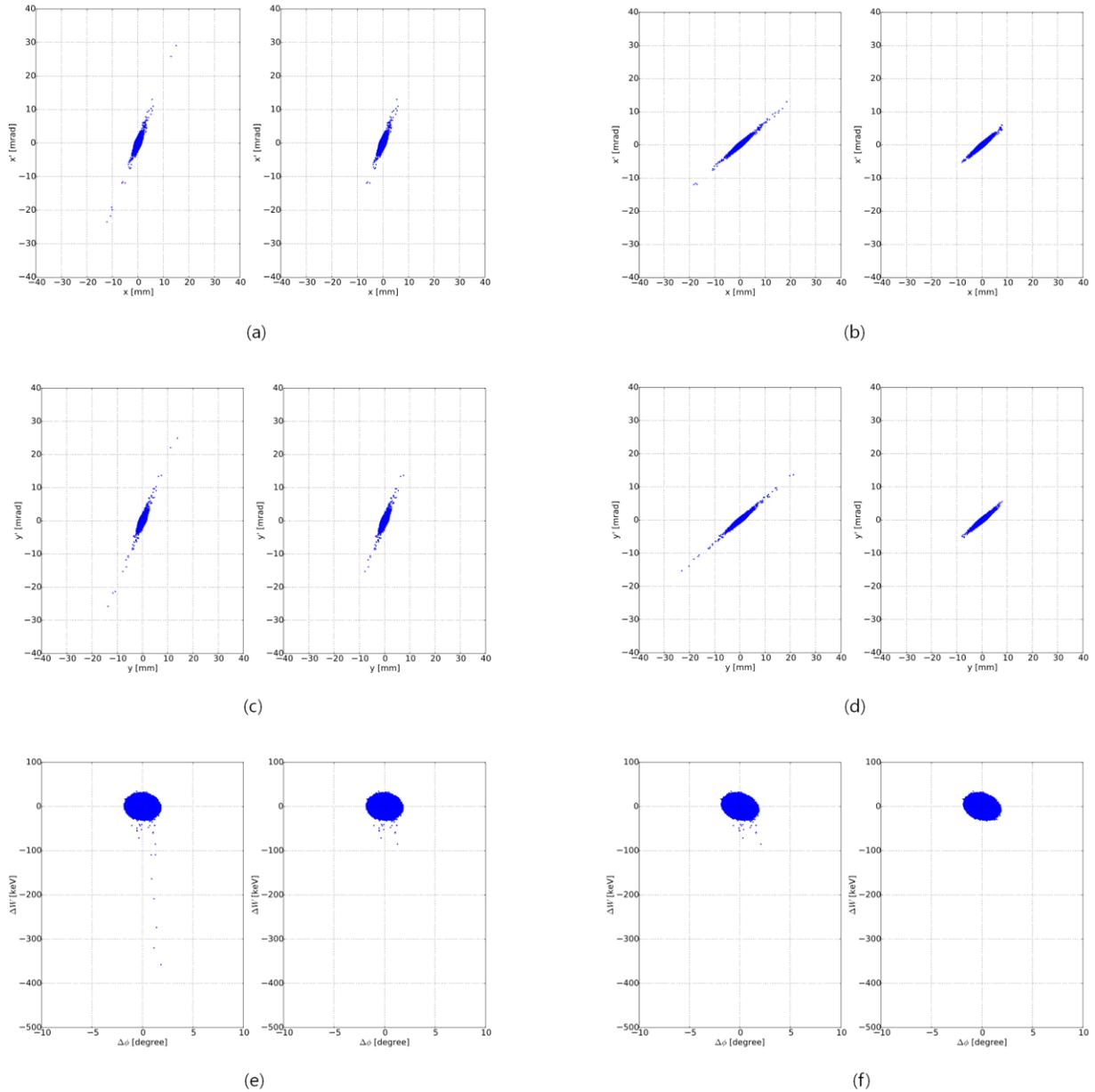

(a)  (b)

(c)  (d)

(e)  (f)

Fig.10. Particle distributions through slits located after the charge stripper in SRIM simulation: (a) x-x' space before and after the first silt, (a) x-x' space before and after the second silt, (c) y-y' space before and after the first silt, (d) y-y' space before and after the second silt, (e) Δϕ-ΔW space before and after the first silt and (f) Δϕ-ΔW space before and after the second silt.

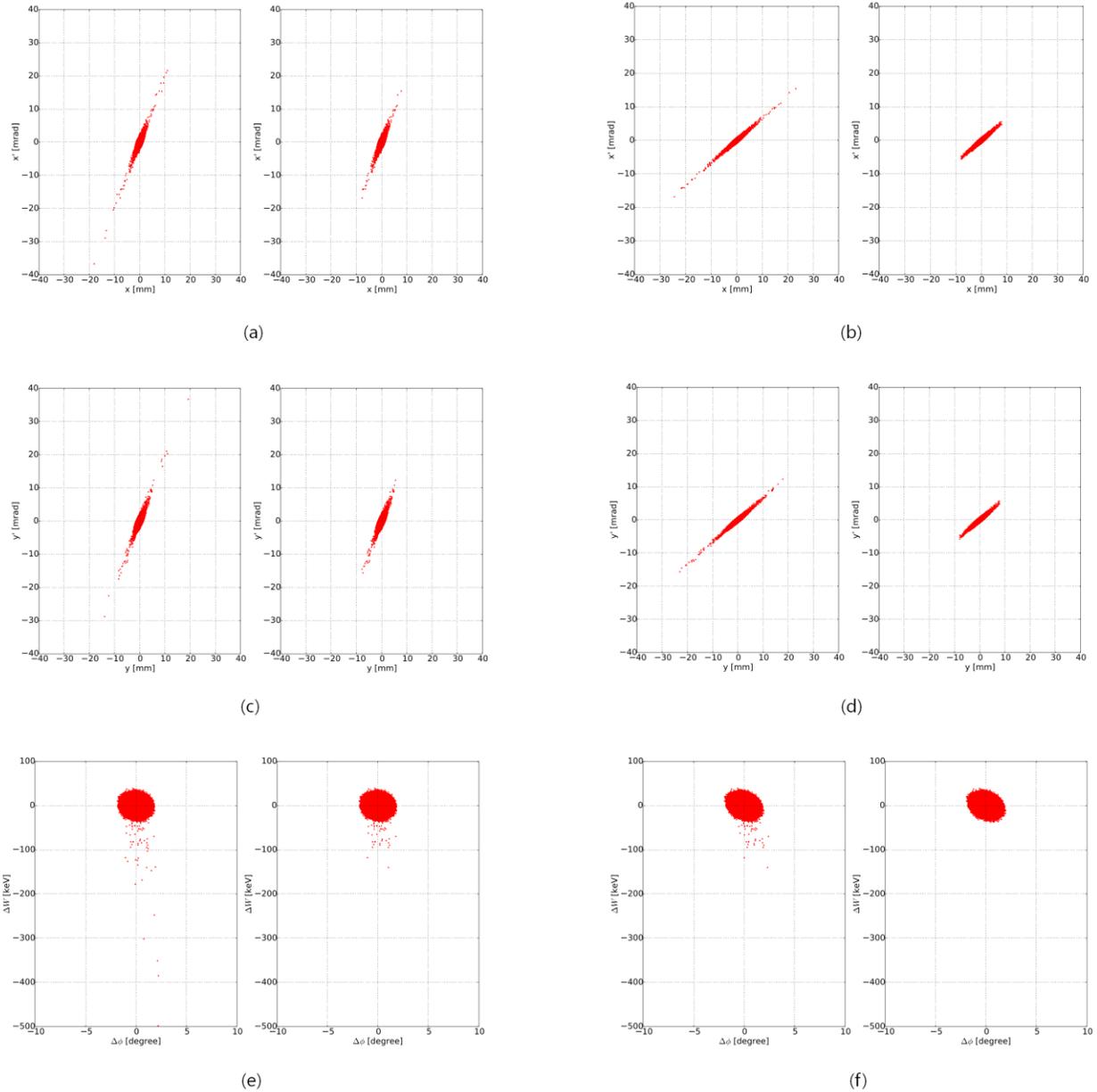

Fig.11. The rms emittances through the 180-degree bending section : (a) in the horizontal direction, (b) in the horizontal direction with smaller scale (c) in the vertical direction and (d) in the longitudinal direction.

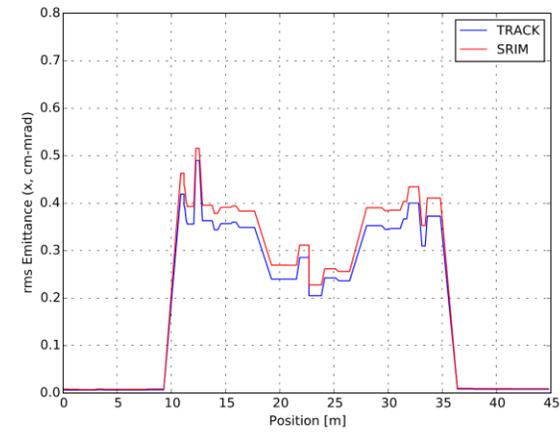

(a)

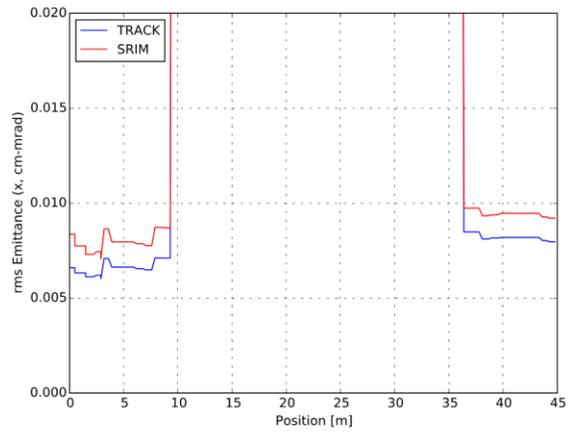

(b)

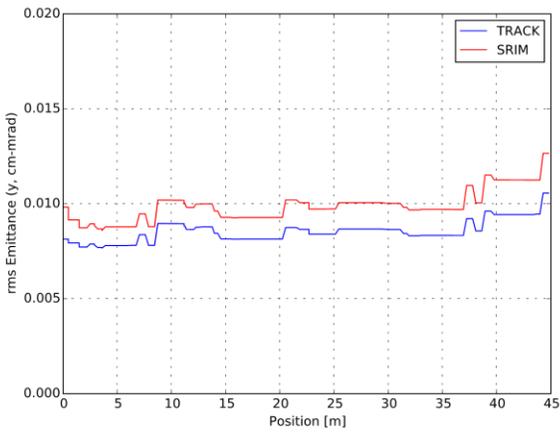

(c)

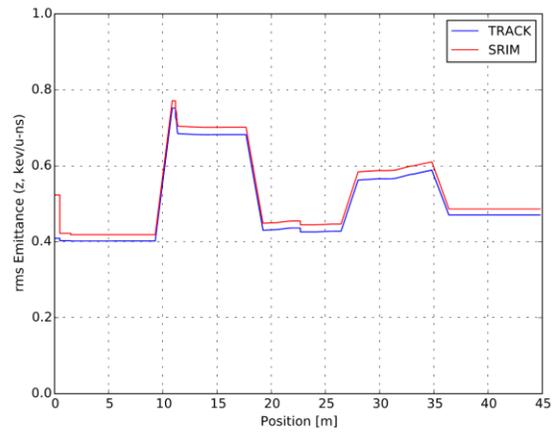

(d)

Fig.12. Particle distribution after the second collimator for charge selection.

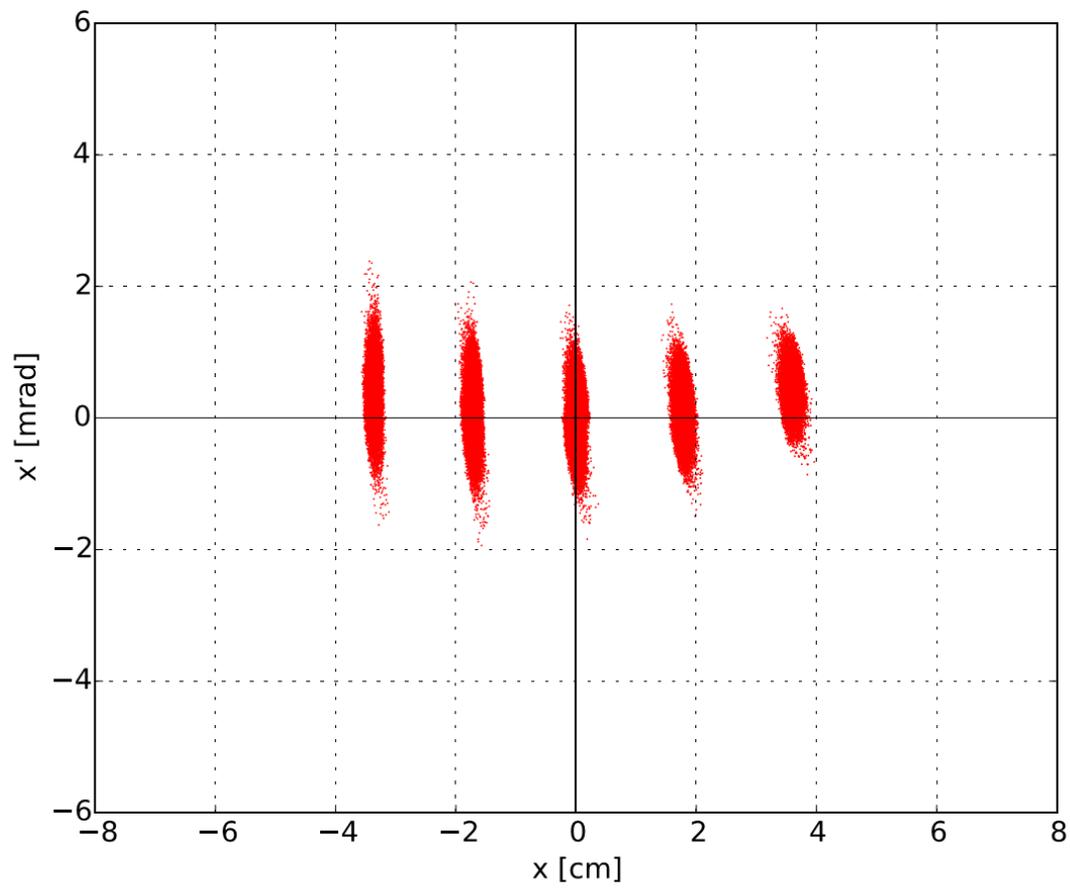